\documentclass[12pt]{article}
\usepackage{amsmath,amssymb}
\usepackage{booktabs}
\usepackage[dvipdfm]{graphicx}
\usepackage{braket}

\begin{document}
  \begin{titlepage}
    \begin{center}
    \hfill TU-868
    ~\\
    \vspace{2cm}
    {\Large\bf
    Tau polarization measurements at the LHC in supersymmetric models
    with a long-lived stau
    }

    \vspace{1cm}
    {\large Ryuichiro Kitano and Mitsutoshi Nakamura}

    \vspace{1cm}
    {\it {Department of Physics, Tohoku University, Sendai 980-8578, Japan}}

    \vspace{3cm}
    \begin{abstract}
      Supersymmetry (SUSY) with a long-lived stau is an attractive
      scenario in the LHC experiments because one can directly observe
      stau tracks in each SUSY event, and thus precise measurements of
      SUSY particle masses are possible. In this scenario, we discuss
      the possibility to observe/measure parity violation in
      interactions among SUSY particles. Such a measurement
      will be important in determining spins and chiralities
      of SUSY particles. We use the last step of the cascade-decay
      chain: $\chi^0 \to \Tilde{\tau}\tau \to \Tilde{\tau} (l \nu \bar \nu)$,
      where the polarization of
      the tau lepton can be determined statistically by looking at
      the energy distribution of the final state lepton.
      Comparing with the theoretical formula of the neutralino
      differential decay width, one can extract the size of parity
      violation in the interaction vertices among the stau, the tau
      lepton and the neutralino. We perform a Monte Carlo simulation to
      see if the effect is visible at the LHC experiments.

    \end{abstract}
    \end{center}
  \end{titlepage}
  \baselineskip 17pt
  
  \section{Introduction}
    If supersymmetry (SUSY) is the solution to the hierarchy problem,
    the superpartners should weigh around a few hundred GeV to TeV
    energy range, which is accessible at the LHC experiments. The way to
    discover SUSY or to measure the properties of SUSY particles are
    quite different depending on the pattern of the superparticle
    spectrum. In particular, the property of the lightest SUSY particle
    (LSP) is important for the search strategies.

    Motivated by an explanation of dark matter of the Universe, many
    collider studies of SUSY models have assumed the case with the LSP
    being the neutralino (admixture of the Higgsinos and the gauginos).
    In the neutralino LSP case, the final states of the SUSY events always
    contain invisible neutralinos, and one needs various measurements to
    be combined in order to determine sparticle masses. The situation
    drastically changes if one assumes that one of the charged sleptons
    is lighter than the lightest neutralino. Such a scenario is possible
    if the slepton is not absolutely stable. For example, the slepton
    can decay into a lepton and gravitino if kinematically allowed.  In
    particular, the scalar tau lepton (stau) can easily be lighter than
    the neutralino in many SUSY models due to the quantum correction
    through the large Yukawa interaction.
 
    If the life-time of the stau is longer than the time scale of the
    collider experiments (a few nsec), we will see charged tracks left
    by staus in each SUSY events, instead of the missing momentum in the
    neutralino LSP scenario.
    Methods to search for a long-lived charged particle in hadron colliders
    have been studied in Refs.~\cite{Drees:1990yw, Feng:1997zr,
    Martin:1998vb, Dimopoulos:1996yq, Nisati:1997gb}.
    Also, various advantages in studying SUSY models at the LHC have
    been reported. The stau momenta and velocities can be measured by
    analyzing the stau tracks, from which the stau mass can be extracted
    with a good accuracy~\cite{Polesello:1999, Ambrosanio:2000ik,
    Ellis:2007}.
    The measurements of other sparticle masses~\cite{Hinchliffe:1998ys,
    Ellis:2006vu, Ibe:2007km, Feng:2009yq, Feng:2009bd, Ito:2009xy} and
    the spin measurement of the stau~\cite{Rajaraman:2007ae} have also
    been discussed.
    Methods to measure the stau life-time have been proposed in various
    contexts~\cite{Buchmuller:2004rq, Hamaguchi:2004df, Feng:2004yi,
    Ishiwata:2008tp, Asai:2009ka}.  The chargino-neutralino production
    and subsequent decay processes at the LHC have been studied in
    Ref.~\cite{Kitano:2008sa} and the possibilities to observe parity/CP
    violations have been discussed.
    
    In this paper, we discuss the possibility to measure the tau
    polarization in the neutralino decay, $\chi^0 \to \tilde \tau \tau$,
    where the neutralinos are mainly produced by the cascade decays of
    colored SUSY particles.
    The tau polarization carries information on whether the lightest
    stau is the partner of the right- or left-handed tau lepton,
    and also  whether the decaying neutralino is mainly composed of 
    the gauginos or the Higgsinos.
    The polarization can be determined by looking at the energy
    distribution of the $\tau$-decay product~\cite{Bullock:1992yt}
    due to the fact that the weak
    interaction maximally violates parity.
    We use the leptonic decay mode of the tau lepton, $\tau \to l \nu
    \bar \nu$, in this work.
    We first show the formula for the lepton-energy distribution
    in the neutralino decay, and by using it we fit the data from the
    Monte Carlo simulation. We find that the size of parity violation
    (the tau polarization) can be measured in a simple model
    where only one of the neutralinos contributes to the final state.
    In a more complicated case, where there are two neutralinos
    contribute to the same final state, we can correctly reproduce
    the signs of parity violation in each decay vertex.

  \section{Neutralino decay}
    In this section, we present the differential decay width of $\chi^0
    \to \Tilde{\tau}\tau \to \Tilde{\tau} (l\bar{\nu_l}\nu_\tau)$ as a
    function of the lepton-energy fraction.  The parity asymmetry of this
    decay process carries information on the chirality of the stau and
    the composition of the neutralino, {\it i.e.,} whether it is
    gaugino-like or Higgsino-like.  

    The measurement of the tau polarization has also been
    studied in the neutralino LSP case.  The effects of the tau
    polarization in the di-tau invariant mass distribution in cascade
    decays were studied in
    Refs.~\cite{Choi:2006mt,Graesser:2008qi,Nattermann:2009gh}.
    Ref.~\cite{Godbole:2008it} studied the $p_T$ distribution of the
    softest $\tau$-jets to extract the tau polarization in the
    co-annihilation region in the mSUGRA model.
   The tau polarization measurements at $e^+e^-$ colliders were studied
   in the stau-pair productions in Refs.~\cite{Nojiri:1994it,
   Bartl:1996wt, Godbole:2004mq}.  The measurements of CP violation at
   $e^+e^-$ colliders were also studied in
   Refs.~\cite{Bartl:2003gr, Choi:2003pq, Dreiner:2010ib}.

  \subsection{Decay width}
    The polarization information of the tau lepton is imprinted to the
    energy distribution of the lepton in the leptonic tau
    decay. We show in this subsection the formula
    for the distribution of the lepton-energy fraction in the neutralino
    decays.

    The relevant interaction Lagrangian for the neutralino decay is
    \begin{align}
      \mathcal{L} = \overline{\chi^0_i}
      ({g_L}_iP_L + {g_R}_iP_R)\tau\Tilde{\tau}^\dagger + {\mbox{H.c.}}~,
      \label{interaction:lagrangian}
    \end{align}
    where $P_L$ and $P_R$ are the chirality projection operators, and
    $g_L$ and $g_R$ are coupling constants.  We will discuss the
    relation to the underlying model parameters in the next subsection.
    The tau lepton is polarized if there is parity violation, {\it i.e.},
    $|g_L| \neq |g_R|$.  The explicit calculation shows that the
    differential decay width of the process, $\chi^0_i \to \tau
    \Tilde{\tau}\to (l\bar{\nu_l}\nu_\tau)\Tilde{\tau}$, is
    \begin{align}
        \frac{
        d\Gamma(\chi^0_i\to\tau \Tilde{\tau} \to 
              l\bar{\nu_l}\nu_\tau\Tilde{\tau})}{\Gamma_\text{lept}}
        = \frac{1}{3} (1-z)\left[
          (5+5z -4z^2) - a_N^{(i)}(1 + z -8z^2) 
          \right]dz~, \label{eq:DifDecayWidth}
    \end{align}
    where $\Gamma_{\rm lept}$ is the partial decay width of the leptonic
    mode~\cite{Bullock:1992yt}.
   The variable $z$ is the lepton-energy fraction:
    \begin{align}
      &z = \frac{E_l}{E_\tau}~,
    \end{align}
    where the energy of the lepton and $\tau$ ($E_l$ and $E_\tau$)
    are those in the rest frame of $\chi^0$.
    The formula (\ref{eq:DifDecayWidth}) is independent of the tau
    charge or the flavor of the final state lepton, {\it i.e.}, electron
    or muon.  The parameter $a_N^{(i)}$ represents the tau polarization
    and it is expressed in terms of the Lagrangian parameters:
    \begin{align}
      a_N^{(i)} = \frac{ |{g_L}_i|^2 - |{g_R}_i|^2 }
                       { |{g_L}_i|^2 + |{g_R}_i|^2 }~.
    \end{align}

\begin{figure}
        \includegraphics[width=8cm,clip]{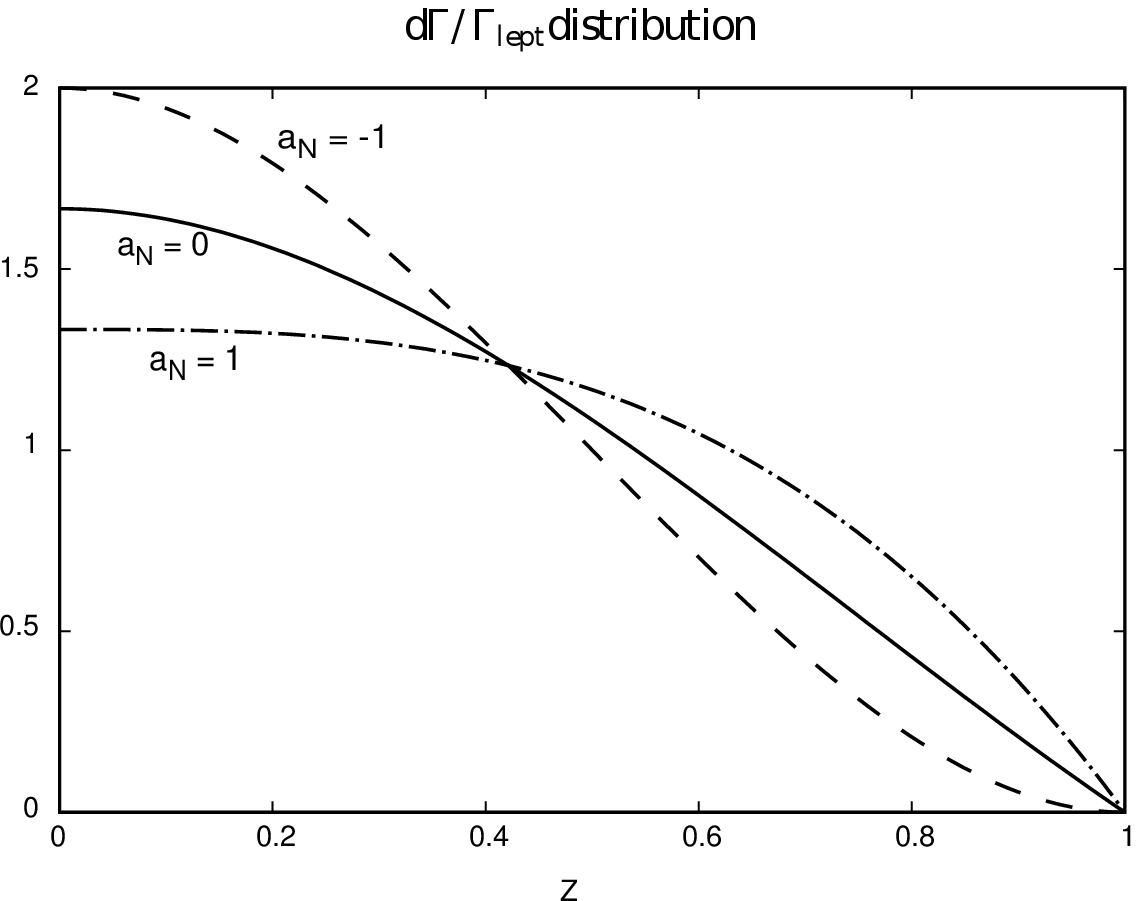}
        \includegraphics[width=8cm,clip]{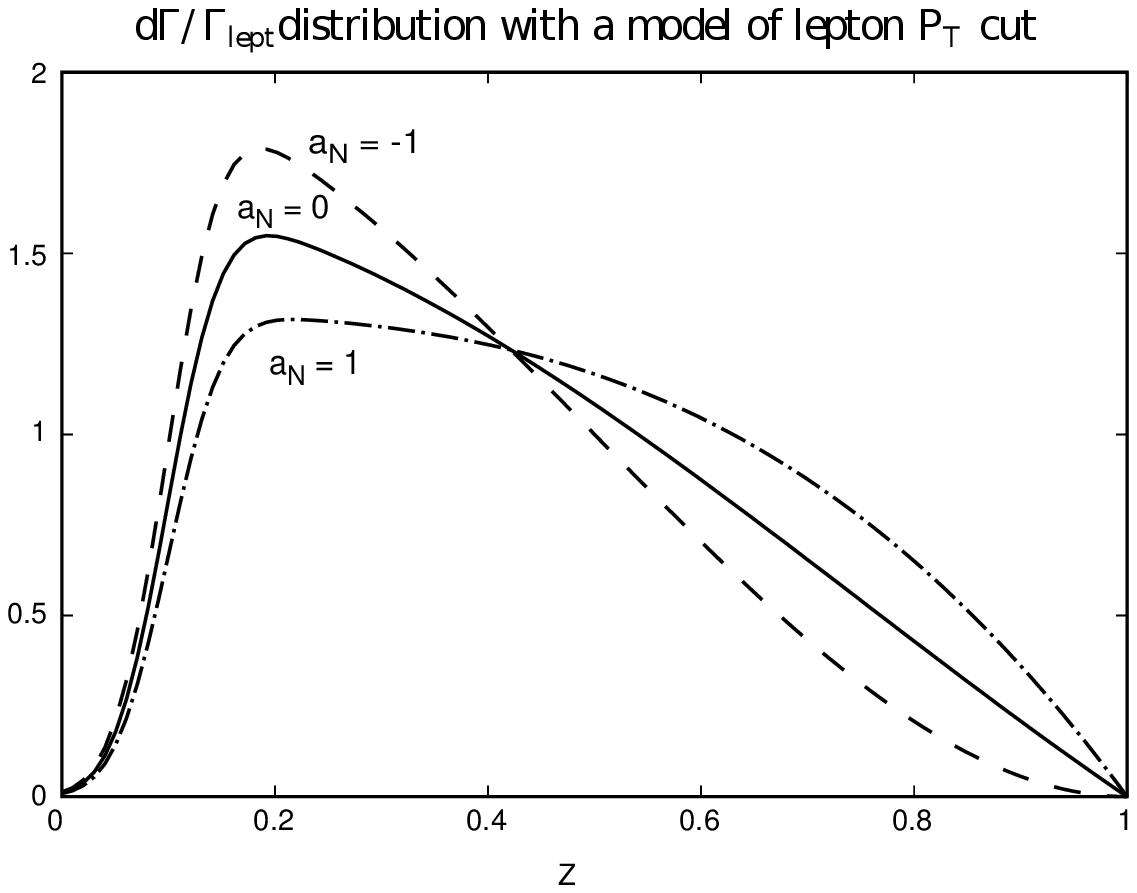}
        \caption{Distributions of the lepton-energy fraction
        ($z=E_l/E_\tau$) in the neutralino decay. The $z$ dependence in
        Eq.~(\ref{eq:DifDecayWidth}) with $a_N = 0,\pm 1$ are shown in
        the left panel.  The right panel is the same figure with a
        simple model to include the effects of the lepton $p_T$ cut.  }
        \label{fig:EnergyDist}
    \end{figure}

    The $z$ distributions in Eq.~(\ref{eq:DifDecayWidth}) for $a_N =
    -1,0 ,+1$ are shown in the left panel of Fig.~\ref{fig:EnergyDist}.
    The value $a_N = -1 (+1) $ corresponds to the decay into right
    (left) handed $\tau$.  The lepton tends to be emitted to the same
    direction of $\tau$ if $a_N = 1$ (left-handed $\tau$).

  \subsection{Coupling constants and the model parameters}
    The coupling constants in Eq.~(\ref{interaction:lagrangian})
    are expressed in terms of the neutralino- and the stau-mixing
    parameters as follows:
    \begin{align}
      {g_L}_i&=\left(-\frac{g_1}{\sqrt{2}} \mathbf{O}_{i1}
               -\frac{g_2}{\sqrt{2}}\mathbf{O}_{i2}
              \right)\sin\theta_{\tilde{\tau}}
          -\frac{m_\tau}{v\cos\beta}\mathbf{O}_{i3}\cos\theta_{\tilde{\tau}}~,\\
      {g_R}_i&=-\frac{m_\tau}{v\cos\beta}\mathbf{O}_{i3}\sin\theta_{\tilde{\tau}} 
          +\sqrt{2}g_1\mathbf{O}_{i1}\cos\theta_{\tilde{\tau}}~.
      \label{coupling:R}
    \end{align}
    where $g_1$ is the coupling constant of the $\rm U(1)_Y$ gauge
    interaction, $g_2$ is that of $\rm SU(2)_L$, $\tan\beta$ is
    the ratio of the two vacuum expectation values (VEV) of the Higgs
    fields and $v$ is the VEV of the Higgs field:
    \begin{align}
      &\tan \beta = \frac{\braket{H_u^0}}{\braket{H_d^0}}~,\\
      &v^2 = \braket{H_u^0} ^2 + \braket{H_d^0}^2~.
    \end{align}
    The matrix elements $\mathbf{O}_{ij}$ are those of a unitary matrix
    which diagonalizes the neutralino mass matrix,
    \begin{align}
      M_{\chi^0} = 
      \begin{pmatrix}
        M_1     &   0   &   -\frac{g_1v}{\sqrt{2}}\cos\beta & \frac{g_1v}{\sqrt{2}}\sin\beta \\
        0       &   M_2 &   \frac{g_2v}{\sqrt{2}}\cos\beta  & -\frac{g_2v}{\sqrt{2}}\sin\beta \\
        -\frac{g_1v}{\sqrt{2}}\cos\beta & \frac{g_2v}{\sqrt{2}}\cos\beta    &   0   &   -\mu  \\
        \frac{g_1v}{\sqrt{2}}\sin\beta  & -\frac{g_2v}{\sqrt{2}}\sin\beta   & -\mu  &     0   
      \end{pmatrix}~,
    \end{align}
    such that
    \begin{align}
      \mathbf{O}M_{\chi^0}\mathbf{O}^T = M_{\chi^0}^{\text{diag}}~.
    \end{align}
    The mixing angle of the stau is defined by
    \begin{align}
      \widetilde{\tau}_1 = \Tilde{\tau}_L\sin\theta_{\tilde{\tau}} 
                         + \Tilde{\tau}_R\cos\theta_{\tilde{\tau}}~.
    \end{align}

    For example, let us consider the case in which the lightest
    neutralino ($\chi^0_1$) and the lightest stau ($\Tilde{\tau}_1$) are
    mainly the Bino and $\tilde \tau_R$, respectively, {\it i.e.,}
    $\cos\theta_{\Tilde\tau} \simeq 1$ and $\mathbf{O}_{11} \gg
    \mathbf{O}_{12}, \mathbf{O}_{13}$.  In this case, the largest
    contribution to the $\chi_1^0$ decay is the second term in
    Eq.~(\ref{coupling:R}). Therefore, the parity violation parameter
    $a_N^{(1)}$ in Eq.~(\ref{eq:DifDecayWidth}) is $a_N^{(1)} \simeq -1$.

    In the case where the decaying neutralino is mainly the Wino and the
    lightest stau is $\tilde \tau_R$, the situation is different.
    Since the pure Wino does not couple to the right-handed stau, the
    decay occurs either through the neutralino mixing or the stau mixing.
    In both cases, the parity violation is $a_N > 0$ in contrast to
    the previous example. 
    
  \section{Monte Carlo simulation}
    In this section, we demonstrate by using a Monte Carlo simulation
    that the distribution obtained in the previous section is observable at the LHC
    experiments.

  \subsection{Basic set up}

    We generated events of SUSY particle productions at a $pp$ collider
    at $\sqrt s = 14$~TeV (LHC) by using the {\tt HERWIG}
    package~\cite{Moretti:2002eu}.
    We use  CTEQ5L~\cite{Lai:1999wy} for the parton distribution
    function.
    The {\tt TAUOLA} package~\cite{Jadach:1993hs} is used for the $\tau$
    decays and the events are passed through the {\tt AcerDET} detector
    simulator~\cite{RichterWas:2002ch} where the lepton momentum are
    smeared to simulate the detector effects.

    The final state contains two stau tracks. The identification of
    those staus can be used to eliminate the background from the
    Standard Model processes.
    In order to distinguish staus from muons, we impose following
    selection cuts on the candidate stau tracks:
      \begin{itemize}
        \item $0.40 < \beta < 0.91$ ,
        \item $p_T > 10 $~GeV ,
        \item $\eta < 2.4 $ .
      \end{itemize}
      Here and hereafter, we do not take into account the momentum and
      the velocity resolutions of the staus, {\it i.e.}, the parton
      level information is used.
    We assume in the following that the stau and neutralino masses are
    known. It has been reported in Refs.~\cite{Ambrosanio:2000ik,
    Ellis:2007, Hinchliffe:1998ys, Ellis:2006vu, Ibe:2007km, Ito:2009xy}
    that those quantities are measurable with a good accuracy.
      
  \subsection{Lepton $p_T$ cut and the deformation of the $z$ distribution}
    We also require $p_T > 15~\rm GeV $ for the lepton momentum.  This
    cut affects the shape of the $z$ distribution in
    Eq.~(\ref{eq:DifDecayWidth}).
    We model the effect of the $p_T$ cut in the $z$ distribution by
    multiplying the following weight:
      \begin{align}
       \frac{1}{2}\, \mathrm{Erfc} 
       \left( \frac{p_T^\text{min} - Cz}{\sqrt{2}\sigma_{p_T}} \right)~,
        \label{eq:PTeffect}
      \end{align}
    where $p_T^\text{min}$ is 15 GeV, and $C$ and $\sigma_{p_T}$ are
    parameters to be determined by fitting the data.  We plot in the
    right panel of Fig.~\ref{fig:EnergyDist} the $z$ distribution
    after multiplying this factor with
    $C=150~\rm GeV$ and $\sigma_{p_T}= 6.0~\rm GeV$.
  
  \subsection{Model I}
    We use the MSSM without flavor mixing and ignore the Yukawa
    interactions of the first and second generations.
    As the first example, we take a simple model in which only one of
    the neutralinos ($\chi^0$) and one of the staus are significantly
    lighter than others, and the lightest neutralino and the
    lightest stau are almost the Bino and right-handed,
    respectively. The value of the parity violation parameter is
		\begin{align}
       a_N = -0.99~, \label{eq:ParityAssmInModelI}
		\end{align}
    in this case.
      
    The MSSM parameters we take are listed in
    Tab.~\ref{tab:Parameter1}. We follow the convention of SUSY Les
    Houches Accord~\cite{Skands:2003cj}.
    The $\chi^0$ and $\Tilde{\tau}$ masses are calculated to be
    \begin{align}
      m_{\chi^0} = 195~\rm{GeV} ~,\ \ \
      m_{\Tilde{\tau}} = 118~\rm{GeV}~.
    \end{align}
    We used the {\tt ISAJET} package~\cite{Paige:2003mg} for the 
    calculation of the mass spectrum and the branching ratios.
    We generate 10,000 SUSY events with the parameter set given in
    Tab.~\ref{tab:Parameter1}.  The number of events corresponds to the
    integrated luminosity of 14.5~$\rm fb^{-1}$.

    \begin{table}
      \begin{center}
        \setlength{\tabcolsep}{3pt}
        \begin{tabular}{c|c}
          \toprule
          \hline
            Parameter      &   Value   \\
          \hline
            $M_1$         &   $195.7$ [GeV]   \\
            $M_2$         &   $800.0$ [GeV]   \\
            $M_3$         &   $1003.7$ [GeV]    \\
            $\tan\beta$   &   $39.46$         \\
            $\mu$         &   $800.0$ [GeV]   \\
            $B$     &   $694 $[GeV]     \\
            $M_{Q_1} ,M_{Q_2} $ & $1400 $[GeV]  \\
            $M_{L_1} ,M_{L_2} $ & $518 $[GeV]   \\
            $M_{Q_3}$     &   $1241 $[GeV]    \\
            $M_{L_3}$     &   $487 $[GeV]     \\
            $M_{u^c}$     &   $1321 $[GeV]    \\
            $M_{d^c}$     &   $1311 $[GeV]    \\
            $M_{c^c}$     &   $1321 $[GeV]    \\
            $M_{s^c}$     &   $1311 $[GeV]    \\
            $M_{e^c}$     &   $278 $[GeV]     \\
            $M_{\mu^c}$   &   $277 $[GeV]     \\
            $M_{b^c}$     &   $1210 $[GeV]    \\
            $M_{t^c}$     &   $1075 $[GeV]    \\
            $M_{\tau^c}$    &   $160 $[GeV]     \\
            $A_t$         &   $-503 $[GeV]    \\
            $A_b$         &   $-567 $[GeV]    \\
            $A_\tau$      &   $-48.3 $[GeV]   \\
          \hline
          \bottomrule
        \end{tabular} 
      \end{center}
      \caption{Parameters of Model~I.}
      \begin{center}
      \parbox{8cm}{}
      \end{center}
      \label{tab:Parameter1}
    \end{table}

    The neutralinos are mainly produced by the cascade decays of the
    squarks and gluinos.
    The value of $z$ (the lepton-energy fraction) in each candidate
    event can be calculated from the measured invariant mass of
    the lepton and the stau,
    \begin{align}
      z = \frac{{M_{\tilde \tau l}}^2 -{m_{\tilde\tau}}^2}
               {{m_{\chi^0}}^2 - {m_{\tilde\tau}}^2}~,
    \end{align}
    where $M_{\tilde \tau l}$ is 
    \begin{align}
      M_{\tilde \tau l}^2 = (p_{\Tilde{\tau}} + p_l)^2~.
    \end{align}
      \begin{figure}
      \centering
      \includegraphics[height=7cm]{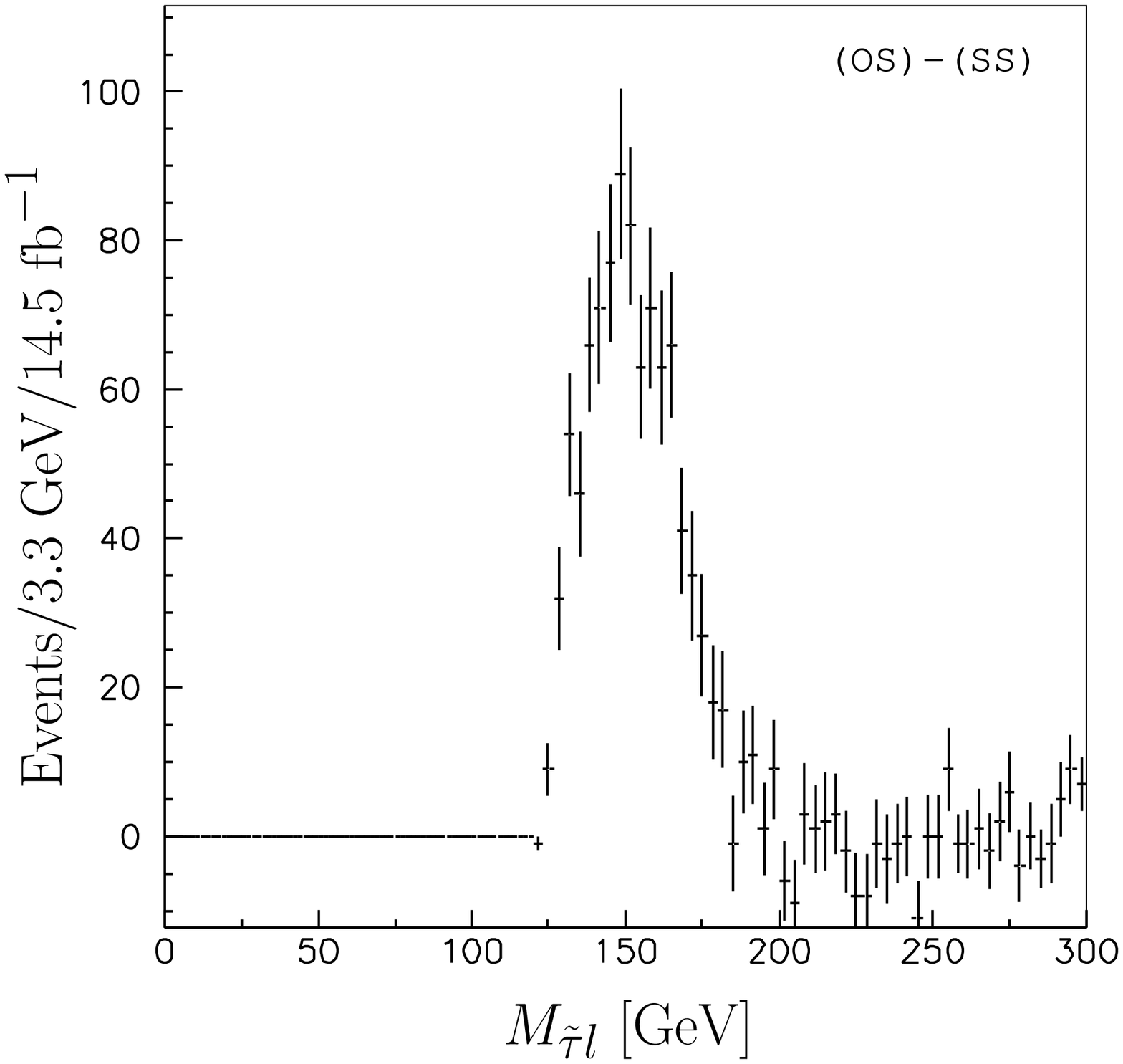}
      \includegraphics[height=7cm]{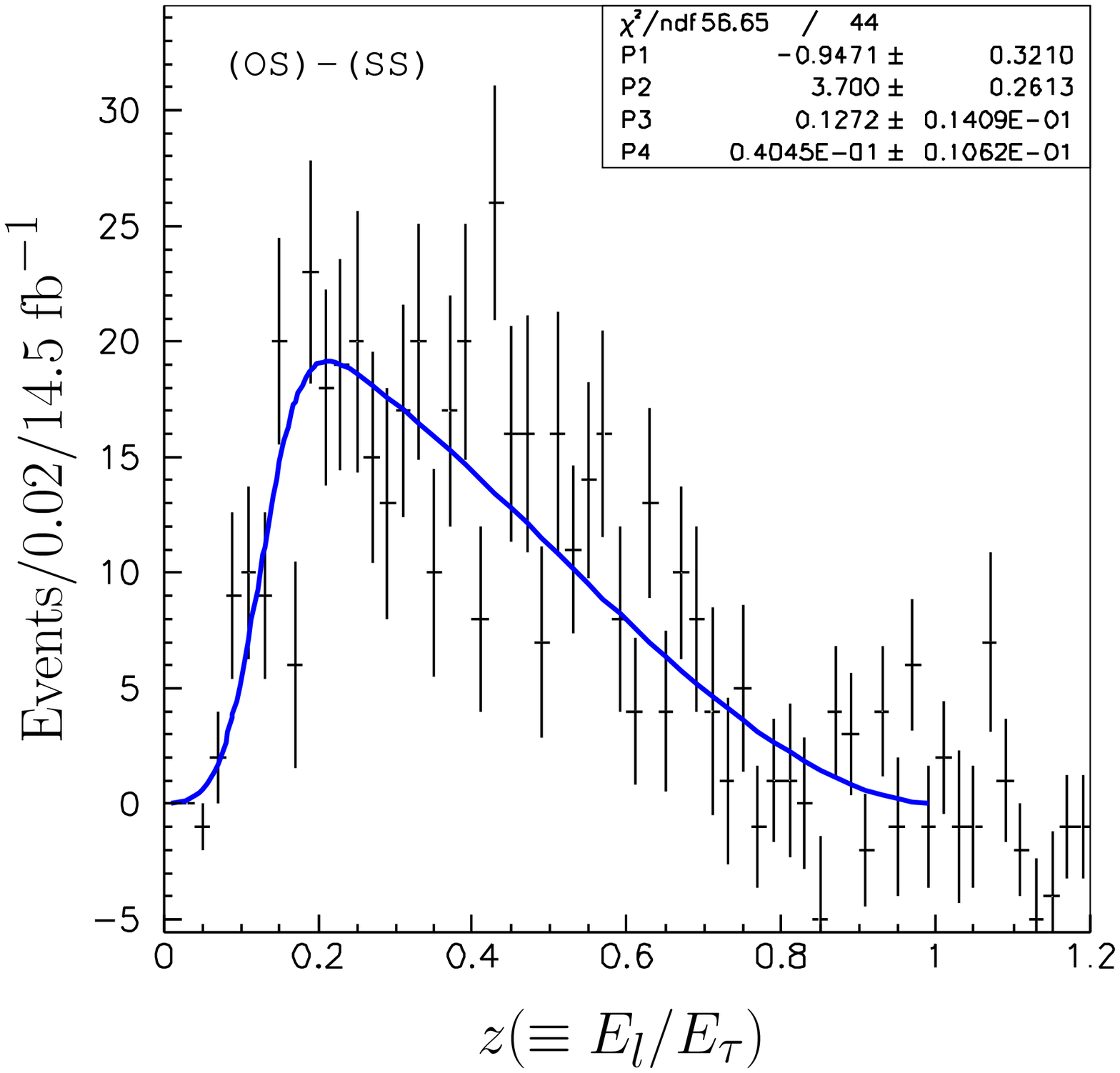}
      \caption{The histograms of the $\Tilde{\tau}$-$l$ invariant mass
        (left) and $z$ ($=E_l / E_\tau$) in Model~I.} \label{fig:Bino}
    \end{figure}
    
    We show in Fig.~\ref{fig:Bino} the histograms of the invariant mass
    (left) and the $z$ parameter (right).  In these plots, we select all
    possible combinations of a stau and a lepton with the opposite
    charges in each candidate event. In order to eliminate the
    contributions from wrong combinations and from background events
    such as leptons from $W$-boson decays, we subtract the same
    distributions calculated by using the combinations with the same
    signs~\cite{Ito:2009xy}. After the selection cuts and the
    charge subtraction explained above, 978 events remained.
  
    By fitting the $z$ distribution in Fig.~\ref{fig:Bino} with the
    function in Eq.~(\ref{eq:DifDecayWidth}) multiplied by a factor in
    Eq.~(\ref{eq:PTeffect}), we obtain the $a_N$ parameter to be:
    \begin{align}
      a_N = -0.95 \pm 0.32 ~.
    \end{align}
    Comparing with the input value in Eq.~(\ref{eq:ParityAssmInModelI}), we can see that the
    parity violation is measured quite successfully.

  \subsection{Model II}
    As the second example, we take a model 
    where the Wino-like neutralino is not so heavy
    compared to the Bino-like one as is so in
    the models motivated by the GUT
    relation of the gaugino masses. The model parameters are listed in
    Tab.~\ref{tab:Parameter2}.
    The neutralino and stau masses are
    \begin{align}
      m_{\chi^0_1} = 195~{\rm GeV}~,\ \ \
      m_{\chi^0_2} = 359~{\rm GeV}~,\ \ \
      m_{\Tilde{\tau}} = 118~{\rm GeV}~.
    \end{align}
    The parity asymmetries $a_N^{(1)}$ and $a_N^{(2)}$ are calculated to be
    \begin{eqnarray}
     a_N^{(1)} = -0.99~, \ \ \ a_N^{(2)} = 1.0~.
    \end{eqnarray}

    Unlike the case of Model~I, we cannot reconstruct $z$ in the
    event-by-event basis since we do not know whether the decaying
    neutralino is $\chi_1^0$ or $\chi_2^0$.  Therefore, we need to
    directly fit the invariant mass distribution which contains events
    of both neutralinos.

    \begin{table}
      \begin{center}
        \setlength{\tabcolsep}{3pt}
        \begin{tabular}{c|c}
          \toprule
          \hline
            Parameter      &   Value   \\
          \hline
            $M_1$         &   $195.7$ [GeV]   \\
            $M_2$         &   $363.9$ [GeV]   \\
            $M_3$         &   $1003.7$ [GeV]    \\
            $\tan\beta$   &   $39.46$         \\
            $\mu$         &   $800.0 $[GeV]   \\
            $B$     &   $694 $[GeV]     \\
            $M_{Q_1} ,M_{Q_2} $ & $1400 $[GeV]  \\
            $M_{L_1} ,M_{L_2} $ & $518 $[GeV]   \\
            $M_{Q_3}$     &   $1241 $[GeV]    \\
            $M_{L_3}$     &   $487 $[GeV]     \\
            $M_{u^c}$     &   $1321 $[GeV]    \\
            $M_{d^c}$     &   $1311 $[GeV]    \\
            $M_{c^c}$     &   $1321 $[GeV]    \\
            $M_{s^c}$     &   $1311 $[GeV]    \\
            $M_{e^c}$     &   $278 $[GeV]     \\
            $M_{\mu^c}$   &   $277 $[GeV]     \\
            $M_{b^c}$     &   $1210 $[GeV]    \\
            $M_{t^c}$     &   $1075 $[GeV]    \\
            $M_{\tau^c}$    &   $160 $[GeV]     \\
            $A_t$         &   $-504 $[GeV]    \\
            $A_b$         &   $-567 $[GeV]    \\
            $A_\tau$      &   $-48.3 $[GeV]   \\
          \hline
          \bottomrule
        \end{tabular}
          \caption{Parameters of Model~II.}
          \label{tab:Parameter2}
      \end{center}
    \end{table}

    \begin{figure}
      \centering
      \includegraphics[height=7cm]{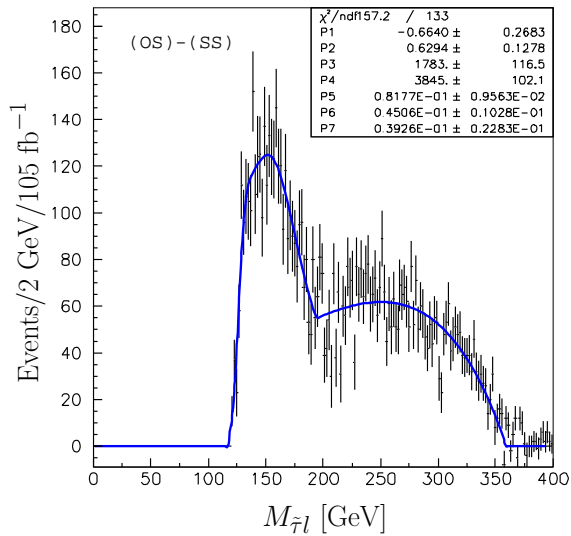}
      \caption{The histogram of the $\Tilde{\tau}$-$l$ invariant
      mass in Model~II.} \label{fig:model2}
    \end{figure}      
    We generated 90,000 SUSY events with this parameter set.  The number
    of events corresponds to the integrated luminosity of 104.6 
    $\rm fb^{-1}$.  We show the invariant mass distributions
    (Fig.~\ref{fig:model2}) using the same technique as we used for the
    Model~I. After the cuts and the subtraction explained before,
    7,677 events remained. We can see contributions from decays of two
    kinds of neutralinos.

    Before fitting the distribution in Fig.~\ref{fig:model2}, we
    checked the validity of Eq.~(\ref{eq:DifDecayWidth}) combined with 
    Eq.~(\ref{eq:PTeffect}) for the Model~II. We
    did it by separating the $\chi_1^0$ and the $\chi_2^0$ events by
    using information from the event generator and performed the
    fitting for each event set. Reasonable values for $a_N^{(1)}$ and
    $a_N^{(2)}$ are obtained:
    \begin{align}
      a_N^{(1)} = -0.80 \pm 0.13~,\ \ \ 
      a_N^{(2)} = 0.94 \pm 0.084~.
    \end{align}

    For the fitting of the histogram in Fig.~\ref{fig:model2}, we
    use the sum of two fitting functions corresponding to each neutralino:
    \begin{align}
      \sum_{i=1,2} N_i \frac{d\Gamma_i(z_i)}{\Gamma^{(i)}_\text{lept}}\times
     \frac{1}{2}\mathrm{Erfc}\left(\frac{p_T^\text{min} -
     C_iz_i}{\sqrt{2}{(\sigma^i_{p_T})}}\right)~,
      \label{eq:fittingBy8}
    \end{align}
    where $p_T^\text{min} = 15~{\rm GeV}$, and 
    $d\Gamma_i/\Gamma^{(i)}_\text{lept}$ is the function in
    Eq.~(\ref{eq:DifDecayWidth}).
   The energy fraction $z_i$ is related to the $\tilde \tau$-$l$
   invariant mass, $M_{\tilde \tau l}$, as
   \begin{eqnarray}
     z_i = \frac{ M_{\tilde \tau l}^2 - m_{\tilde \tau}^2 }
                { m_{\chi_i^0}^2 - m_{\tilde \tau}^2 }~.
   \end{eqnarray}
   There are eight fitting parameters $a_N^{(i)}$, $N_i$, $C_i$ and
   $\sigma^i_{p_T}$ for $i = 1,2$ in the function.

    Since it is difficult to find the best fit for such many parameters,
    we here make a (semi) theoretic assumption on a relationship between
    two parameters $C_1$ and $C_2$ in order to reduce the number of
    parameters.
    In Eq.~(\ref{eq:fittingBy8}), the combinations of $C_i z_i$
    represent the average values of $p_T^l$ from the $\chi_i^0$ decay
    in the laboratory frame.
    Since the neutralinos are produced mainly from heavy colored
    particles, it is likely to be highly boosted and thus the tau lepton
    from the neutralino decay is pointing to the similar direction to
    the one of the neutralino in the laboratory frame. Therefore, there
    is an approximate relation:
    \begin{eqnarray}
     \langle p_T^{\text{lept}} \rangle
      = z_i \langle p_T^{\tau} \rangle
      \simeq z_i  \Braket{\frac{ E_\tau }{ E_{\chi^0_i}} }
      \langle p_T^{\chi^0_i} \rangle~,
    \end{eqnarray}
    where $\braket{{E_\tau} / {E_{\chi^0_i}}}$ is
    \begin{align}
      \Braket{\frac{E_\tau}{E_{\chi^0_i}}} = \frac{m^2_{\chi^0_i} -
     m^2_{\Tilde{\tau}}}{2m_{\chi^0_i}^2}~.
    \end{align}
    The average transverse momentum of the neutralino,
    $\braket{p_T^{\chi^0_i}}$, is independent of the neutralino mass if
    it is highly boosted. Therefore, the ratio of $C_1/C_2$ is
    estimated to be
    \begin{align}
      \frac{C_1}{C_2} =
        \frac{ m^2_{\chi^0_2} }{ m^2_{\chi^0_1} }
        \frac{ m_{\chi^0_1}^2 - m^2_{\Tilde{\tau}} }
             { m^2_{\chi^0_2}-m_{\Tilde{\tau}}^2   } ~.
    \end{align}
    By imposing the above relation, we could fit with the 
    function in Eq.~(\ref{eq:fittingBy8}):
    \begin{align}
      a_N^{(1)} =  -0.66 \pm 0.27 ~,\ \ \
      a_N^{(2)} = 0.63 \pm 0.13 ~.
    \end{align}
    We obtain the correct signs for each parity violation
    although the magnitude of the parity violation is obtained to be
    smaller than the theoretical input values ($-0.99$ and $1.0$,
    respectively) due to our simple ansatz on the fitting function.

  \section{Summary}

    We studied the polarization measurement of the tau lepton at the LHC
    in the long-lived stau scenario.  The polarization of the tau lepton
    from the neutralino decays carries information on whether the
    neutralino is Higgsino-like or gaugino-like and the chirality of the
    partner of the stau.
    
    We have shown that the $\tau$ polarization can be measured by
    fitting the distribution of the lepton-energy fraction 
    ($E_l /E_\tau$) in the leptonic $\tau$ decays. We performed Monte Carlo
    simulations for two parameter sets of the MSSM. The first
    example we took is a simple case where the Bino-like neutralino is
    significantly lighter than others. In this case, we can directly fit
    the energy fraction by the theoretical function and the parity
    violation can be measured successfully.

    In the second case where the Wino-like neutralino also contributes
    to the same final state, we fit the $\tilde \tau$-$l$ invariant mass
    distribution with the two contributions summed. Although the fitting
    gives a milder asymmetry compared to the theoretical inputs, we can
    obtain qualitatively correct features of parity violation in each
    neutralino decay.

\section*{Acknowledgements}

    We would like to thank T.~Moroi and T.~Ito for useful discussions. RK is
    supported in part by the Grant-in-Aid for Scientific Research
    21840006 of JSPS.

\end{document}